\documentclass[aps,prc,onecolumn]{revtex4}
\usepackage{epsfig}
\newcommand{\be}{\begin{equation}} 
\newcommand{\ee}{\end{equation}}
\newcommand{\bea}{\begin{eqnarray}} 
\newcommand{\eea}{\end{eqnarray}}

\newcommand{\gton}{\mathrel{\lower.9ex \hbox{$\stackrel{\displaystyle 
>}{\sim}$}}} 
\newcommand{\lton}{\mathrel{\lower.9ex \hbox{$\stackrel{\displaystyle 
<}{\sim}$}}}

\newcommand{\vn}{{{\bf n}}} 
\newcommand{\vnull}{{{\bf 0}}} 
\newcommand{\vp}{{{\bf p}}}

\newcommand{\vv}{{\bf v}}

\begin{document}
\title{Generating random thermal momenta}
\author{Denes Molnar}
\affiliation{Physics Department, Purdue University,
525 Northwestern Avenue,
West Lafayette, IN 47907, USA}

\date{\today}
\begin{abstract}

\end{abstract}

%\pacs{12.38.Mh, 25.75.-q, 25.75.Ld}
\maketitle
Generation of random thermal particle momenta is a basic task in many
problems, such as microscopic studies
of equilibrium and transport properties of systems, or the conversion of
a fluid to particles. In heavy-ion physics, 
the (in)efficiency of the algorithm matters particularly
in hybrid {\em hydrodynamics + hadronic transport} 
calculations. With popular software packages, such as 
{\sf UrQMD 3.3p1}~\cite{hybridUrQMD} or 
{\sf THERMINATOR}~\cite{therminator}, it
can still take ten hours to generate particles for a single Pb+Pb ``event''
at the LHC from fluid dynamics output.
Below I describe reasonably efficient basic 
algorithms using the {\sf MPC} package
(Molnar's Program Collection) \cite{MPC}, which should help speed momentum
generation up by at least one order of magnitude.

It is likely that this wheel has been reinvented many times instead of reuse, 
so there may very well exist older and/or better algorithms that I am not
aware of ({\sf MPC} has been around only since 2000). 
The main goal here is to encourage practitioners to use
available efficient routines, and offer a few practical solutions.

%%%
\subsection{Static thermal distributions}

First consider a static thermal distribution
\be
f_0(\vp) \equiv \frac{dN}{d^3p} 
  = \frac{g}{(2\pi)^3} \frac{1}{e^{(E(\vp) - \mu)/T} + a} \qquad, 
\qquad\qquad E(\vp) = \sqrt{\vp^2 + m^2}
\label{static_dist}
\ee
where $\mu$ is the chemical potential, $m$ is the particle mass, $g$ is the 
degeneracy factor (internal degrees of freedom), and 
\be
a = \left\{
      \matrix{
         +1 & {\rm for\ fermions} \hfill \cr
         -1 & {\rm for\ bosons} \hfill \cr
         0 & {\rm for\ classical\ (Boltzmann)\ particles} \cr
             }
    \right.
\ee
The distribution is isotropic, 
so in standard spherical coordinates
we have {\em uniformly} distributed
\be
\varphi \in [0,2\pi) \quad \ , \qquad \cos \vartheta \in [-1,1] 
\ee
and the only challenge is to generate the scaled magnitude $x = |\vp|/T$
\be
g(x) \equiv \frac{dN}{dx} \propto \frac{x^2}{e^{\sqrt{x^2 + z^2}} + \alpha} 
\quad\qquad\qquad \qquad 
(z \equiv \frac{m}{T}\ , \quad \alpha \equiv a\, e^{\mu/T})
\ee

\begin{figure}
\leavevmode
\epsfysize=6cm
\epsfbox{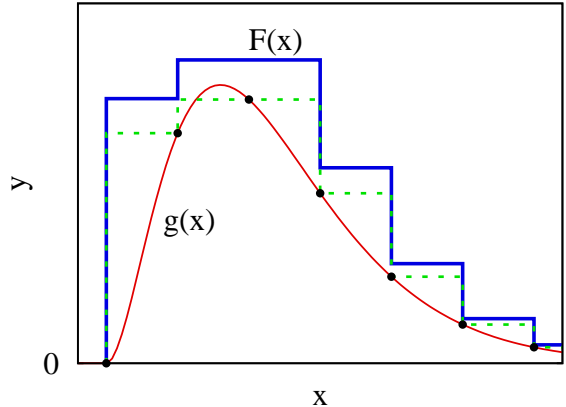}
\caption{Illustration of rejection method with staircase 
comparison function.}
\label{Fig:rejection}
\end{figure}
A straightforward technique to employ is an automated 
version of the {\em rejection method}
(see Fig.~\ref{Fig:rejection}). The rejection method\cite{rejection}
is based on a suitable comparison 
function $F$ that bounds $g$ from above for all $x$.
One then generates uniformly
$(x,y)$ pairs in the two-dimensional 
area under the graph of $F$, and keeps the $x$ 
coordinate of those points that are below the graph of $g$.
It is easy to construct a {\em staircase} comparison function,
automatically,
via tabulating $g(x)$ at several points, and multiplying by a modest
``safety'' factor
to leave room for any local maxima within the intervals
\cite{autorejection}.
Uniform sampling
under the graph of a staircase function is straightforward%
\footnote{
E.g., generate a uniform deviate $I$, 
and find the interval $[x_i,x_{i+1}]$
that has the point $x$ satisfying 
$I = P(x) \equiv \int\limits_0^x dx' F(x') / \int\limits_0^{\infty} dx' F(x')$.
With precalculated $P(x_i)$ this boils
down to a binary search. Then pick $y$ uniformly in $[0,F(x)]$.
}\cite{autorejection}.

Alternatively, we can tabulate $g(x)$, construct an 
{\em interpolation} of it, and 
generate the interpolated distribution%
\footnote{
See {\sf TabulatedRndDist1D} class in {\sf MPC}.
}. Computationally this is very similar to staircase function sampling in 
automated rejection. To precisely
represent $g(x)$, in general many points are needed but we then have $100$\%
efficiency (there is no rejection check).

%%%
\subsection{Boosted thermal distributions}

For thermal distribution with collective flow
\be
f_v(\vp) \equiv \frac{dN}{d^3p} 
  = \frac{g}{(2\pi)^3} \frac{1}{e^{(p^\alpha u_\alpha - \mu)/T} + a} \quad, 
\qquad p^0 = \sqrt{\vp^2 + m^2} \quad, \qquad u^\mu = \gamma(1, \vv_F)
\label{boosted_dist}
\ee
where $\vv_F$ is the three-velocity of the fluid. This is 
a special case ($n^\mu = (1, \vnull)$)
of Cooper-Frye freezeout discussed in the next Section.

%%%
\subsection{Cooper-Frye freezeout}

In Cooper-Frye freezeout\cite{CooperFrye} 
a fluid is converted to particles across a 
three-dimensional spacetime hypersurface. A surface element $d\sigma^\mu(x)$
at spacetime point $x$ contributes to particle number
\be
dN = p^\alpha d\sigma_\alpha(x) \frac{d^3p}{p^0}\, f(x,\vp)  \qquad
\qquad \qquad (d\sigma_\alpha \equiv n_\alpha d^3 V \ , 
\quad n^\alpha = (n^0, \vn) \ , 
\quad n^2 = \pm 1)
\ee
where $f$ is the phase space density of the particles in the fluid. 
For an ideal fluid $f$ is thermal, so
the momentum distribution of particles emitted from the surface element is
\be
f_{CF}(\vp) = \frac{dN}{d^3p} \propto 
\frac{p_\alpha n^\alpha}{p^0} f_v(\vp)
\label{CF_dist}
\ee
with $f_v$ from (\ref{boosted_dist}).
Because $dN$ is a Lorentz scalar, in the local rest frame of the
fluid element ($\tilde \vv_F = \vnull$) we have a static thermal distribution
with a momentum-dependent prefactor
\be
\frac{dN}{d^3 \tilde p} \propto \frac{\tilde p_\alpha \tilde n^\alpha}{\tilde p^0} f_0(\tilde \vp) = 
(\tilde n^0 - \tilde \vn \tilde \vv) f_0(\tilde \vp)\ , \qquad\qquad \tilde \vv \equiv \frac{\tilde \vp}{\tilde p^0}
\ee
After generating a thermal momentum $\tilde \vp$, the prefactor 
can be accounted for in an additional rejection step.
If $\tilde \vp$ fails the rejection test, we generate a new one,
otherwise we boost it back to the original frame to obtain $\vp$.

For {\em timelike} normal with $\tilde n^0 > 0$, the prefactor is always 
nonnegative,
and the additional rejection step is
at least 50\% efficient because we can use as comparison function
$\tilde n^0 + |\tilde \vn| < 2 \tilde n^0$, while
on average $\langle \tilde n^0 - \tilde \vn\tilde \vv \rangle = \tilde n^0$. Negative 
$\tilde n^0 < 0$ is unphysical ($dN$ is never positive then). So we need
no more than {\em two tries}, on average.

For {\em spacelike} normal ($n^2 = -1$), there are always momenta for which 
$(\tilde n^0 - \tilde \vn \tilde \vv)$ is positive, and also momenta
for
which it is negative, so one needs to follow some prescription to handle
unphysical contributions.
A customary choice is to ignore negative
values and implement a {\em cut} prefactor
$(\tilde n^0 - \tilde \vn \tilde \vv) 
\Theta(\tilde n^0 - \tilde \vn \tilde \vv)$ instead,
even if with thermal $f_0$ conservation laws are violated%
\footnote{
Unless we also alter the hydrodynamic variables that define the distribution.
See, e.g., 
%\cite{Anderlik:1998cb}
%\bibitem{Anderlik:1998cb} 
%C.~Anderlik, Z.~I.~Lazar, V.~K.~Magas, L.~P.~Csernai, H.~Stoecker and W.~Greiner,
C.~Anderlik {\it et al},
  %``Nonideal particle distributions from kinetic freezeout models,''
  Phys.\ Rev.\ C {\bf 59}, 388 (1999)
  [nucl-th/9808024].
  %%CITATION = NUCL-TH/9808024;%%
}.
For $\tilde n^0 \ge 0$, 
a comparison function $\tilde n^0 + |\tilde \vn|$
then provides%
\footnote{
If $\tilde n^0 \ge |\tilde \vn| |\tilde \vv|$, the $\Theta$-function is unity 
and the efficiency has the same expression as for a timelike normal
$\tilde n^0 /(\tilde n^0 + |\tilde \vn|) \ge |\tilde \vv| /2$
because $|\vn| > \tilde n^0$.
While if $0 \le \tilde n^0 < |\tilde \vn| |\tilde \vv|$, 
on average $\langle (\tilde n^0 - \tilde \vn\tilde \vv)\Theta(\tilde n^0 - \tilde \vn\tilde \vv) 
\rangle = (\tilde n^0 + |\tilde \vn| |\tilde \vv|)^2 / (4 |\tilde \vn| |\tilde \vv|) \ge (\tilde n^0 + |\tilde \vn| |\tilde \vv|)  
(\tilde n^0 + |\tilde \vn|) / (4|\tilde \vn|)$, so
the efficiency is at least 
$(\tilde n^0 + |\tilde \vn||\tilde\vv|) / (4|\tilde \vn|) \ge |\tilde \vv|/4$.
}
 an efficiency of at least $\langle |\tilde \vv|\rangle / 4$,
which for $m/T \gg 1$ is poor but still workable in heavy-ion
physics applications
because spacelike surface elements are rare, and heavy particles have
low abundances.
For $\tilde n^0 < 0$ the total particle
yield from the hypersurface element (without the cut) is negative, 
and for that reason such surface elements
are most often omitted.
Momentum generation 
for $\tilde n^0 < 0$, if desired, requires additional
care%
\footnote{When $|\tilde \vn|$ is close to $|\tilde n^0|$,
only high-velocity particles moving in nearly opposite direction to 
$\tilde \vn$ can contribute because otherwise the prefactor is negative,
and a naive rejection algorithm can turn into a near-infinite loop.
For reasonable efficiency one must generate {\em directly}
only the allowed tail and limited solid angle 
of the static thermal distribution $f_0$ or of 
distribution (\ref{modified_f0}). 
The constraints
are i) $|\tilde \vv| \ge |\tilde n^0|/|\tilde \vn|$, i.e., 
$|\tilde \vp| \ge m |\tilde n^0|$; 
and ii) the angle between $\tilde \vn$ and $\tilde \vp$
satisfies $\cos\tilde \vartheta \le -|\tilde n^0|/|\tilde \vn|$.}.

For better efficiency,
especially for spacelike normals and heavy particles,
we can generate directly the isotropic 
\be
(\tilde n^0 + |\tilde \vn| |\tilde\vv|)\,
 \Theta(\tilde n^0 + |\tilde\vn| |\tilde\vv|)\, 
f_0(\tilde \vp) \ ,
\label{modified_f0}
\ee and realize the correct prefactor
 $(\tilde n^0 - \tilde\vn \tilde \vv) \Theta(\tilde n^0 - \tilde\vn \tilde \vv)$
via rejection using the more efficient%
\footnote{
In particular, for spacelike normals with $\tilde n^0 \ge 0$, the
efficiency is 
$\langle (\tilde n^0 - \tilde \vn \tilde \vv) 
\Theta(\tilde n^0 - \tilde \vn \tilde \vv) \rangle  / (\tilde n^0 + |\tilde \vn| |\tilde \vv|) =
(\tilde n^0 + |\tilde \vn| |\tilde \vv|) / (4 |\tilde \vn| |\tilde \vv|)
\ge 1/4$.
} comparison function
$\tilde n^0 + |\tilde \vn| |\tilde\vv|$.
This way,
for {\em timelike} normals
we need no more than {\em two tries}, on average,
while for {\em spacelike} normals with $\tilde n^0 \ge 0$, 
no more than {\em four tries}.
The price is that a new generator has to be constructed
for each hypersurface element, 
whereas the static thermal generator ($f_0$) can be reused when 
$T=const$ across the entire hypersurface (often the case in practice).

%%%
\subsection{Massless Boltzmann distribution}

Finally, I show that for {\em massless} Boltzmann particles
static thermal momenta 
\be
\frac{dN}{dx} \propto x^2 e^{-x}  \quad , 
\qquad\qquad\qquad x \equiv \frac{|\vp|}{T} 
\ ,
\ee
can be generated through the one-liner%
\footnote{Routine {\sf boltzmandev0} in {\sf MPC}.}
\be
x = -\ln[(1-\xi_1)(1-\xi_2)(1-\xi_3)]
\ee
where $\xi_{1,2,3}$ are three independent uniform deviates on $[0,1)$.
This follows because with $x = r^2$, the distribution corresponds to
a Gaussian in six-dimensional space 
$\{y_1, y_2, \dots, y_6\}$
\be
\frac{dN}{r^5 dr} \propto e^{-r^2}  
\qquad \ , \qquad\qquad r = \sqrt{y_1^2+y_2^2+ \cdots + y_6^2}
\ee
which can be generated as three independent {\em pairs} of 
Gaussian deviates, each pair through 
the standard 2D polar coordinate technique
$dx\, dy\, e^{-x^2}\,\! e^{-y^2} 
\propto d\varphi\, d(\rho^2)\, e^{-\rho^2}$. Here
$\varphi$ is uniform on $[0,2\pi)$ and $\rho^2$ is exponential on 
$[0,\infty)$. We only need $\rho^2$ for each pair
because $r^2 = \rho_{12}^2 + \rho_{34}^2 + \rho_{56}^2$, and exponential deviates
can be generated through inverting
\be
I(\rho^2) = \int\limits_0^{\rho^2} dw\, e^{-w} = 1 - e^{-\rho^2} 
\qquad\qquad \Rightarrow\qquad \rho^2 = -\ln(1 - I)
\ee
where $I$ is a uniform deviate on $[0,1)$.

%%%
\subsection{Conclusion}

In a paradigm where random numbers are computationally very expensive,
a useful efficiency benchmark for three-momentum generators is
\be
\epsilon = \frac{3}{\langle uniform\ deviates\ per\ momentum \rangle}
\ee
(i.e., with one random number generator call per
momentum component, $\epsilon = 1$).
For the algorithms above -
static thermal (Sec. A), boosted thermal (Sec. B), and 
Cooper-Frye (Sec. C) - 
$\epsilon = 1$, $\epsilon \ge 3/8$, and $\epsilon \ge 3/16$, respectively,
if we use the interpolation method%
\footnote{
With automated rejection the formal 
efficiency drops to $\epsilon \gton 3/4$, 
$\epsilon \gton 3/10$, $\epsilon \gton 3/20$ because with a 
nearly perfect comparison function we still need 
two random deviates per call.
}.

In reality there is of course 
overhead due to function tabulation, memory lookups, 
and binary search, which limit
 the number of intervals one can utilize. 
But in practice, 
overall efficiency is still a reasonable $\sim 0.1$
(even with the automated rejection method),
and often much better%
\footnote{The main inefficiency in {\sf UrQMD} and {\sf THERMINATOR} 
comes from the
use of a {\em constant} comparison function, directly for the
three-dimensional $f(\vp)$.}.

%%%
\medskip
{\bf Acknowledgments. -} Discussions with Giorgio Torrieri, Pasi Huovinen, and
Sangwook Ryu are greatfully acknowledged.
The routines described here were developed under DOE grants 
DE-FG-02-93ER-40764 (Columbia), DE-FG02-01ER41190 (Ohio State),
DE-AC02-98CH10886 (RIKEN BNL), and DE-PS02-09ER41665 (Purdue).

\end{document}